\documentclass[a4paper]{article}

\usepackage{INTERSPEECH2022}
\usepackage{hyperref}

\title{Expressive Singing Synthesis Using Local Style Token \\ and Dual-path Pitch Encoder}
\name{Juheon Lee$^{1,2}$, Hyeong-Seok Choi$^{1,2}$, Kyogu Lee$^{1,2}$}
\address{
  $^1$Music and Audio Research Group (MARG), Seoul National University\\
  $^2$supertone Inc.}
\email{ [juheon2, kekepa15, kglee]@snu.ac.kr}

\begin{document}
\ninept

\maketitle
\begin{abstract}
This paper proposes a controllable singing voice synthesis system capable of generating expressive singing voice with two novel methodologies. First, a local style token module, which predicts frame-level style tokens from an input pitch and text sequence, is proposed to allow the singing voice system to control musical expression often unspecified in sheet music (e.g., breathing and intensity). Second, we propose a dual-path pitch encoder with a choice of two different pitch inputs: MIDI pitch sequence or f0 contour. Because the initial generation of a singing voice is usually executed by taking a MIDI pitch sequence, one can later extract an f0 contour from the generated singing voice and modify the f0 contour to a finer level as desired. Through quantitative and qualitative evaluations, we confirmed that the proposed model could control various musical expressions while not sacrificing the sound quality of the singing voice synthesis system.
\end{abstract}


\section{Introduction}

Singing voice synthesis (SVS) is the task of generating a natural singing voice from a given musical score.
With the development of various deep generative models, research on synthesizing high-quality singing voice has been emerging recently \cite{chen2020hifisinger, zhuang2021litesing, hono2021sinsy, liu2021diffsinger}.
As the performance of the SVS improves, there are increasing cases in which the technology is applied to the production of actual music content \cite{huang2020ai}.
Accordingly, the SVS system that can control various musical expressions by reflecting the user's intention draws more attention.

There are two challenging problems in building an SVS system that can easily control various expressions. 
First, it is challenging to build datasets where various musical expressions are annotated. 
Unlike information such as pitch and lyrics, which are relatively easy to label, expressive elements such as breathing, intensity, and singing techniques related to pitch contour are more expensive and time-consuming to label. 
The second problem is that the more information the user has to enter into the SVS system initially, the more burdensome it will become to the user.
In general, the SVS system takes a MIDI pitch sequence and lyrics as an input.
Although lots of input parameters such as musical expressions (e.g., breath, intensity, vibrato parameters) can be given to the system, it is inconvenient for the user because the number of input parameters one has to specify at frame-level increases whenever creating a new song.
Therefore, to build an SVS system that is easy to use and can control various expressions while dealing with these problems, 1) expression elements should be trained based on self-supervised manners, and 2) the parameters for the detail expressions should be generated automatically during the initial generation stage but should still be able to be modified and resynthesized if desired.

To this end, we propose an SVS system capable of controlling expression along with two novel methods. 
First, to model a variety of unlabeled style representations, we introduce a Local Style Token (LST) module that captures styles in a self-supervised manner from given text and pitch information based on \cite{wang2018style}.
Unlike \cite{wang2018style}, however, we do not design the model to infer a single global style vector from a reference signal but rather predict frame-wise style tokens that change over time. 

Second, to take control over the f0 contour, we introduce a Dual-path Pitch Encoder (DPE) that can selectively use MIDI pitch and f0 contour as input. 
In the training process, the output of the two encoders is randomly selected to produce the same result. In the generation process, we can resynthesize the singing by freely controlling the f0 contour extracted from the results generated by the MIDI pitch.
Through the quantitative and qualitative evaluation, we confirmed that the proposed system allows free control of expressions such as breathing, intensity, and detailed f0 techniques while producing a high-quality singing voice.

\begin{figure*}[ht]
\centering
\includegraphics[width=0.95\linewidth]{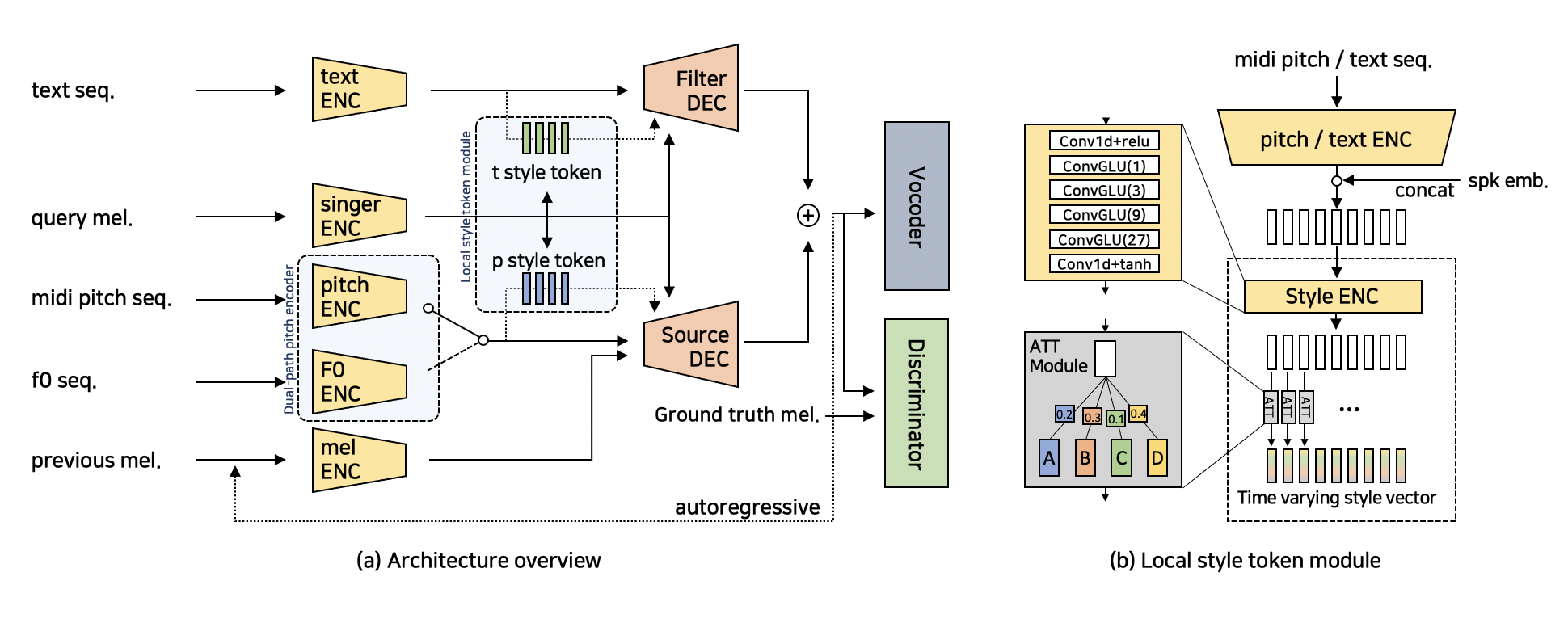}
\caption{The proposed expressive SVS system overall structure diagram (left) and local style token module diagram (right). $ConvGLU(k)$ denotes 1D-CNN layer with gated linear unit in which kernel size is $k$.} 
\label{fig:overview}
\end{figure*}

The main contributions of this study are as follows:

\begin{itemize}
    \item We propose a content-driven local style token module that can model various musical expressions in singing, such as intensity and breathing, trained in a self-supervised manner.
    \item We propose a dual-path pitch encoder that takes either MIDI pitch sequence or f0 contour, allowing the users to control pitch at a coarse or fine level of one's choice.
\end{itemize}

\vspace{-1em}

\section{Proposed System}

Our proposed SVS model is designed by adding a local style token (LST) and dual-path pitch encoder (DPE) to model and control various expressions based on \cite{lee2020disentangling}. 
Based on the source-filter theory \cite{chiba1958vowel}, the acoustic model of our SVS system is an auto-regressive model including two decoders that generate the filter and the source signal, respectively. 
The filter and source were designed to be modeled from (text) and (pitch, previous acoustic feature), respectively, where singer embedding and LST are used together as conditions. Specifically, acoustic feature $\hat{M}$ is generated as follows:
\begin{equation}
 \hat{M} = D_{F}(E_{t}, cat(E_{s}, T_t)) + D_{S}(E_{m}, E_{p}, cat(E_{s}, T_p)),
\end{equation}

where $D_F, D_S$ denotes filter and source decoder, respectively. $E_{t/s/m}$ denotes encoder output of each input, and $T_{t/p}$ means local style token sequence. 
Finally, the generated acoustic feature $\hat{M}$ is converted to a waveform through a vocoder. The entire network is trained with acoustic feature reconstruction loss and adversarial loss, and the overview of the acoustic model structure is shown in Fig. 1.

\subsection{Local style token module}
We introduce a local style token module to model unspecified singing expressions in a music score. We assume that musical expression elements in singing can be inferred from a given input text and pitch sequence and that these elements should exist in a time-varying form. 
To achieve this goal, we modify the attention mechanism proposed in \cite{wang2018style} and retrieve a local style token sequence by referencing the input contents such as pitch and text.
We first introduce a style encoder consisting of stacked 1d-CNN layers with gated linear units \cite{dauphin2017language} to obtain query sequence $Q_{t}, Q_{p} \in \mathbb{R}^{L \times d}$ from text $E_t$, pitch $E_p$ and singer $E_s$ embedding sequences as follows:
\begin{equation}
    Q_{t/p} = \mathrm{StyleEnc}(cat(E_{t/p}, E_s))
\end{equation}
where $L, d$ denotes sequence length and channel dimension, respectively, and $E_s$ is obtained by broadcasting the singer encoder output vector as much as length $L$. 
Because the LST module operates exactly the same on both text $t$ and pitch $p$ sides, we denote the subscript $t$ or $p$ as $t/p$ throughout the paper.

Then, $N$ randomly initialized trainable style key and value $K_{t/p}, V_{t/p} \in \mathbb{R}^{N \times d}$ is used to obtain the style score $S_{t/p} \in \mathbb{R}^{L \times N}$, and is computed as follows:

\begin{equation}
    {S}_{t/p} = \mathrm{softmax}(\frac{Q_{t/p} K_{t/p}^T}{\sqrt{d}})
\end{equation}
Finally, we obtain a LST sequence via matrix multiplication between $S_{t/p}$ and $V_{t/p}$ as follows:  $T_{t/p} = S_{t/p} V_{t/p}$.
In the inference stage, the predicted LST sequence from the input contents may be used as it is, or the style score $S_{t/p}$ can be modified in the desired way in frame-level to control the musical expression of the singing voice, as shown in Fig. 2.
The overview of the LST module is illustrated in Fig. 1-(b).

\subsection{Dual-path pitch encoder}
Another essential factor that determines the expressiveness of singing is the f0 contour. Modeling a natural f0 contour from a MIDI pitch sequence is one of the important research areas of SVS research \cite{ohishi2012stochastic, wada2018sequential, kameoka2013generative}. 
However, the natural f0 contour must be carefully determined by referencing not only MIDI pitch information but also text, singer information \cite{lee2012study, ikemiya2014transferring}. 

Meanwhile, the model proposed in \cite{lee2020disentangling} produces a spectrogram having natural f0 implicitly reflecting information from inputs to the system such as singers, MIDI pitch sequence, and lyrics.
Inspired by this, we aim to design a model that can use both MIDI pitch and f0 contour as inputs instead of making an additional model that predicts f0 contour explicitly.
To this end, a dual-path pitch encoder with the same structure in which two inputs of pitch and f0 contour can be freely used is proposed, and the training is conducted by randomly selecting one of the two pitch representations.
This way, we can generate singing voice with natural f0 contour from the initial generation using MIDI pitch input.
If we want further control pitch techniques such as vibrato or portamento, we can modify f0 directly and recreate them using the f0 encoder, as shown in Fig. 2.

\subsection{Bandwidth extension vocoder}
We used a HiFiGAN vocoder \cite{kong2020hifi} to convert the generated acoustic feature into a waveform. 
Interestingly, we found that the HiFiGAN vocoder can simultaneously perform bandwidth extension and waveform generation. 
That is, we trained the vocoder to convert a 22.05khz acoustic feature generated by the acoustic model into a 44.1khz waveform to generate a higher quality sound source without having to train an acoustic model that generates a 44.1khz acoustic feature.

\begin{figure}[ht]
\centering
\includegraphics[width=1.0\linewidth]{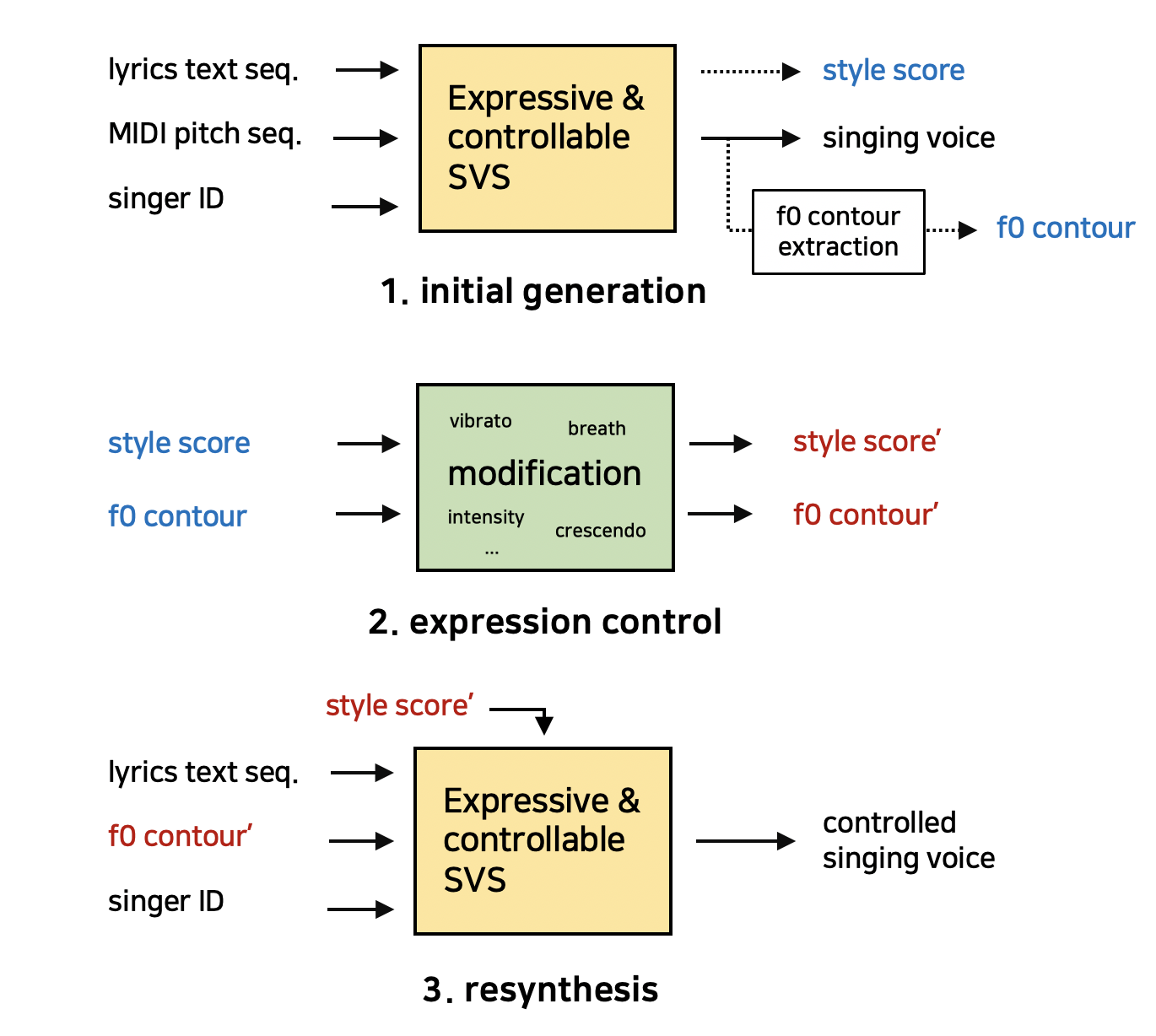}
\caption{The control procedure of the proposed SVS system. First, the singing voice is generated using the music score as an initial generation step. We obtain the style score and f0 contour from the initially generated singing voice. Next, the style score and f0 contour are modified as desired by the user. Finally, we obtain the desired singing voice by resynthesizing it with the modified inputs.}
\label{fig:overview2}
\end{figure}

\vspace{-1em}

\section{EXPERIMENT}

\subsection{Dataset}
We used an internal dataset of 1,150 singing voices of 88 females and 66 males for training.
Each singing voice is paired with manually annotated musical scores.
The sampling rate of the singing voices was set to 44.1khz, and the notes were annotated for each syllable.
A phoneme-level annotation of lyrics was done by assigning one frame to onset and coda each, and the rest of the frame to vowel as proposed in \cite{lee2019adversarially}.

\subsection{Training}
The training of the models was done the same as in \cite{lee2020disentangling} except for the newly proposed LST and DPE.
The number of the style tokens $N$ was set to 4, and training was conducted using either MIDI pitch sequence or f0 contour with a 50\% probability for each pitch input. 
We used WORLD \cite{morise2016world} to extract f0 contour from the singing voices. The encoder and decoder structure of the model is the same as \cite{lee2020disentangling} except that all highway convolutional units\cite{srivastava2015training} have been changed to GLUs. The structure of the style encoder is as shown in Fig. 1-(b), and the structure of the f0 encoder is the same as that of the MIDI pitch encoder except for the input channel of the first 1d-CNN layer.
We used a 128-dimensional mel spectrogram extracted from the waveform of the 22.05kHz sampling rate with window size and hop length set to 1024, 256, respectively, as an acoustic feature. 
The acoustic model was trained using the adversarial loss proposed by \cite{lee2019adversarially} along with L1 loss.

We trained three models for comparative experiments to see if LST and DPE help improve controllability while generating high-quality singing voices. The three models are as follows: 
1) \textbf{Single} model has only MIDI pitch encoder without an f0 encoder. The f0 contour is controlled by modifying the initially generated singing voice with WORLD vocoder \cite{morise2016world}.
2) \textbf{Dual} has a DPE including both an f0 and a MIDI pitch sequence encoder. 
3) \textbf{DualLST} model incorporates both the DPE and LST modules.

\subsection{Quantitative evaluation}
To examine the generation quality and controllability of the proposed model, we organized two test sets for listening evaluations.
The first test set consisted of singing voices generated with MIDI pitch and text input. Ten male and ten female singers were randomly selected to generate ten musical verses. 
To account for degradation from vocoders, we reconstructed the waveforms from ground truth acoustic features using the vocoders and included them in the listening evaluations.
Secondly, we conducted a pitch shift experiment to verify if it is possible to control the f0 contour extracted from the initially generated results.
We randomly selected 60 musical scores and initially generated them using the midi pitch input. Then, after raising or lowering each of the 60 f0 contours extracted from the generation result by two semitones, re-synthesis was performed to generate pitch shifted sound sources.
The listening evaluations were conducted through Amazon Mechanical Turk, and 10 participants per sound source evaluated the overall naturalness and pitch naturalness of each sound source. The evaluation results for each test are shown in Table 1 and Table 2, respectively. \footnote{Audio sample : \url{https://tinyurl.com/cpyfbt6h}}

\begin{table}[h]

\caption{ Initial generation test MOS  result }

\centering
\scalebox{0.9}{
\setlength\tabcolsep{7pt}
\begin{tabular}{l|c|c}

\textbf{Model}      & \textbf{overall naturalness} & \textbf{pitch naturalness}  \\ \hline

Single     & $3.86 \pm 0.05$ & $3.89 \pm 0.05$ \\
Dual     & $3.79 \pm 0.06$ & $3.86 \pm 0.06$ \\
DualLST     & $3.84 \pm 0.05$ & $3.87 \pm 0.05$ \\
\hline
Recon     & $3.89 \pm 0.05$ & $3.90 \pm 0.06$ \\
GT     & $4.03 \pm 0.05$ & $4.06 \pm 0.05$ \\

\end{tabular}
}
\end{table}

\begin{table}[h]

\caption{ Pitch shift test  MOS result }

\centering
\scalebox{0.9}{
\setlength\tabcolsep{7pt}
\begin{tabular}{l|c|c|c}

\textbf{Model}      & \textbf{-2} & \textbf{0} & \textbf{+2}  \\ \hline
Single     & $3.75 \pm 0.18$ & $3.69 \pm 0.18$ & $3.63 \pm 0.22$\\
Dual     & $3.80 \pm 0.16$ & $3.82 \pm 0.17$ & $3.80 \pm 0.19$\\
DualLST     & $3.80 \pm 0.18$ & $3.92 \pm 0.15$ & $3.78 \pm 0.16$\\

\end{tabular}
}
\end{table}

Table 1 shows that all of the models obtained naturalness results that did not differ significantly from reconstructed singing voices by the vocoder.
Using DPE resulted in a slightly lower naturalness score than the Single model, which seems to be the result of the ambiguity that occurred in producing the same result from two different types of inputs. However, this difference was not statistically significant, so we claim that the DPE model can control the f0 contour while maintaining the naturalness similar to the baseline model.

Table 2 shows that using DPE can still produce high-quality sound sources even when the pitch is shifted than using the parametric vocoder capable of f0 control. 
The Dual and DualLST models showed better naturalness than the Single model in all cases. Although we performed the global pitch shift in the experiment to maintain the temporal context of the song, the pitch shift can be applied to local sections. The model we proposed can naturally reflect various pitch techniques such as vibrato, attack, and release, as introduced in 3.5.

\begin{figure*}[t]
\centering
\includegraphics[width=0.95\linewidth]{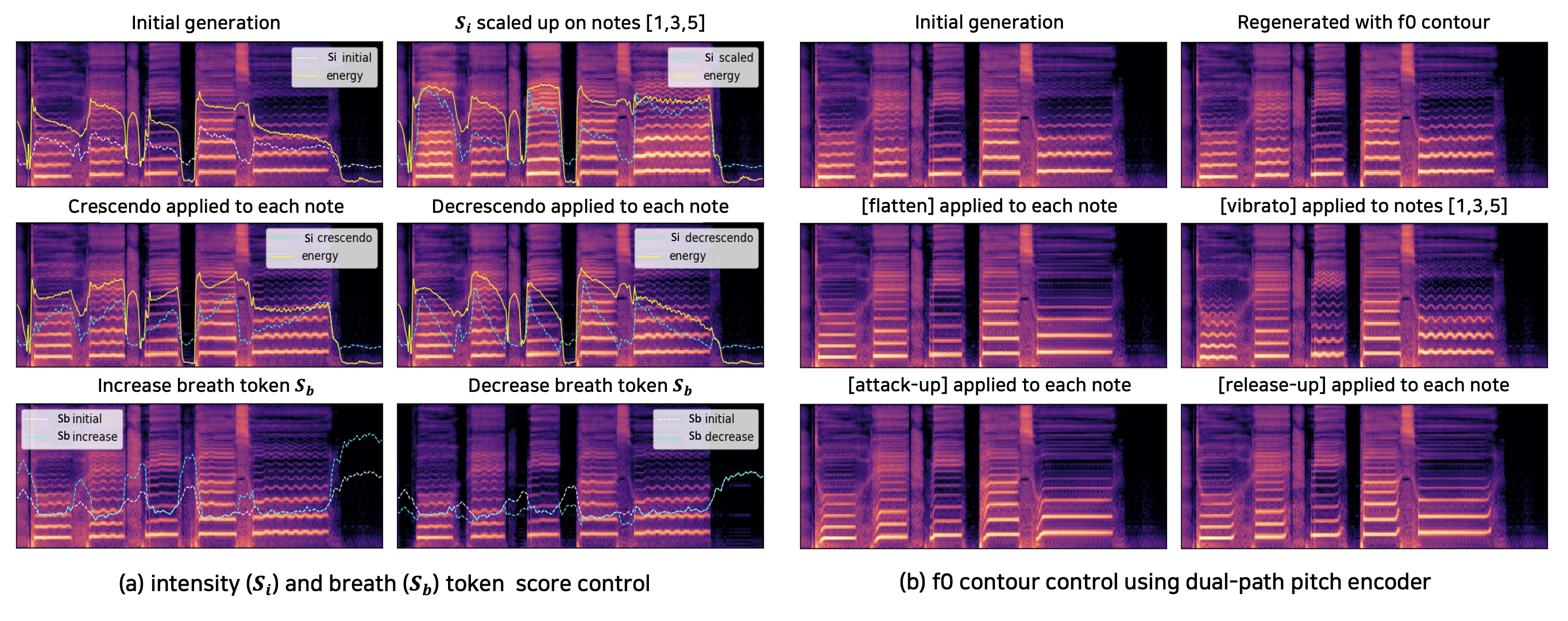}
\caption{Spectrogram analysis with (a) intensity, breath and (b) f0 contour control applied.}
\label{fig:spectrogram analysis}
\end{figure*}

\subsection{Dual-path reconstruction analysis}
A reconstruction analysis was performed to quantitatively confirm if DPE helps faithfully reflect the input pitch information.
First, we initially generated 150 audio samples from randomly selected singers and phrases. Then, we extracted f0 contours from the generated samples and reconstructed them using the extracted f0 contours.
Finally, we measured Mel Cepstral Distortion (MCD), f0-RMSE, and V/UV error rates between the initially generated and the reconstructed samples.
The results in Table 3 show that the difference between the initially generated sample and the reconstructed sample is small, showing that the proposed DPE module is working as intended.
We also report the error between the ground truth audio sample and the reconstructed sample using a HiFiGAN vocoder, which is shown as \textit{Recon} in Table 3.
This shows that the difference between the initially generated and the reconstructed sample is sufficiently small even when compared to the difference between the ground truth audio sample and the reconstructed sample using the vocoder.
In particular, adding the LST module always helped lowering evaluation measures. It implies that the LST module not only helps control singing expressions but is also helpful for generating better output with accurate f0 contour.

\begin{table}[h]
\caption{ MCD, f0-RMSE and V/UV for reconstruction test }
\centering
\scalebox{0.9}{
\setlength\tabcolsep{7pt}
\begin{tabular}{l|c|c|c}
\textbf{Model}  & \textbf{MCD(dB)} & \textbf{f0 RMSE(Hz)} & \textbf{VUV(\%)}  \\ \hline
Dual     & $3.03$ & $7.21$ & $3.04$ \\
DualLST  & $2.95$ & $7.06$ & $2.93$ \\
\hline
Recon & $4.69$ & $ 5.61$ &  $3.54$\\ 
\end{tabular}
}
\end{table}

\subsection{Qualitative analysis}
To qualitatively examine the controllability of the proposed methods, we tried various style modifications by manipulating the initial LST sequence and f0 contour \footnote{The role of each of the $N$ style tokens is randomly permuted for every experiment. In addition, although breath and intensity were always captured by the style tokens in the text side ($T_t$), no meaningful token was found in the pitch side ($T_p$).}.
\\
\textbf{Breath control $\mid$} 
The first noticeable style captured by the style tokens was the token activated in the breathing section of the phrases.
We named it as breath token ($V_{b} \in V_t$).
By changing the style score ($S_{b} \in S_t$) of the breath token from half to double, we found that we can easily control the intensity of breathing sounds, as shown in Fig. 3-(a)-bottom. 
\\
\textbf{Intensity control $\mid$} We also found that one of the tokens captures the intensity of the singing voice. We named it an intensity token ($V_i \in V_t$). Here, the term \textit{intensity} was used as a word to mean the energy and the timbre of singing according to the vocalization method, such as falsetto, chest voice.
As shown in Fig. 3-(a)-top, we found that we can change the intensity of the singing voice, which is shown explicitly by an energy contour.
Taking advantage of this, we found that we can control musical expressions such as crescendo or decrescendo by linearly increasing or decreasing the attention score of the intensity token ($S_{i} \in S_t$).
\\
\textbf{f0 contour control $\mid$} We can easily control f0 contour using DPE with simple operations such as reducing variance (flatten), adding sinusoidal values (vibrato), adding or subtracting small values from the onset and offset positions of the note (attack/release, up/down), as shown in Fig. 3-(b). Note that, compared to directly adjusting the quantized midi pitch, f0 contour control makes it possible to reflect more detailed expressions in Hz unit.

\section{Related work}

Recently, interest in research on the SVS system that can reflect musical expression is increasing.
A method of explicitly modeling information such as pitch curves, energy, V/UV., which can be extracted directly from the vocal signal, was proposed in \cite{zhuang2021litesing}. \cite{hono2021sinsy} proposed a method to interpret the music score more naturally by introducing a module that predicts the difference between the actual singing and the score. Efforts to create natural pitch contour have also been made in various ways, such as directly predicting f0 from note sequences \cite{wada2018sequential, ohishi2012stochastic, lee2012study}, or predicting variables of the parametric f0 contours \cite{bonada2020hybrid}.
Despite various kinds of efforts to improve the expressive power of the SVS system, there has been no study yet that, to our knowledge, allows users' to control singing style elements that cannot be extracted directly from the signal.

\section{conclusion}

We proposed a local style token module and a dual-path pitch encoder to design an SVS system capable of modeling and controlling various musical expressions. We confirmed that the LST token predicted from contents can be controlled to modify expressions such as intensity and breathing. The f0 contour can be controlled through DPE to express various singing techniques related to pitch control. Listening evaluations showed that the proposed model could generate a high-quality singing voice by reflecting the users' intentions.

\bibliographystyle{IEEEtran}

\bibliography{mybib}


\end{document}